\documentclass[10pt,letterpaper]{article}

\usepackage{cogsci}
\usepackage{pslatex}
\usepackage{apacite}
\usepackage{float} 
\usepackage{graphicx}
\usepackage{booktabs}   
\usepackage{multirow}  
\usepackage[table]{xcolor}
\usepackage{xcolor}
\usepackage{colortbl}    
\usepackage{todonotes}
\usepackage{mathptmx}
\usepackage[ruled,vlined]{algorithm2e}
\usepackage{enumitem}
\usepackage{amsmath}
\usepackage{threeparttable}

\newlist{scales}{description}{1}
\setlist[scales]{
  style=nextline,
  leftmargin=2em,
  labelsep=0.5em,
  itemsep=1pt,
  topsep=2pt
}

\title{Measuring Cognitive Engagement in Collaborative Discourse with an Extended ICAP Framework: Comparing Human Annotation, In-Context Learning, and Reflective LLM Agents}

\author{
Lan Anh Do$^{1,*}$ (lan\_anh.do@tufts.edu) 
Hanling Jiang$^{2,*}$ (hanling.jiang@tufts.edu) \\
Shuchin Aeron$^{2}$ (shuchin.aeron@tufts.edu) 
Ayanna K.~Thomas$^{1}$ (ayanna.thomas@tufts.edu) \\
$^{1}$Department of Psychology, Tufts University \\
$^{2}$Department of Electrical and Computer Engineering, Tufts University \\
$^{*}$These authors contributed equally as co-first authors.
}

\begin{document}

\maketitle

\begin{abstract}
Collaboration supports learning and problem-solving, but its effectiveness depends on cognitive engagement during discourse. This study applies an extended 7-point ICAP framework based on the Interactive, Constructive, Active, and Passive modes to characterize variation in cognitive engagement during collaborative dialogue. Engagement was coded by trained human annotators and compared with large language model (LLM)–based labeling approaches, including in-context learning (ICL), zero-shot prompting, and self-reflective agents. Interrater reliability among human annotators was robust across framework refinement stages ($\kappa = 0.906$–$0.998$), higher than the moderate agreement observed for ICL-based annotation ($\kappa = 0.541$–$0.609$). The human-refined framework improved agreement among human annotators ($\Delta\kappa = 0.10$), but produced only modest gains for ICL-based LLMs ($\Delta\kappa < 0.04$). Agent-refined frameworks improved cross-model agreement but remained below the human-refined framework. These findings highlight the promise of agent-based approaches and the importance of continued interaction between theory-guided human annotation and LLM-based methods in future work.
 
\end{abstract}

\section{Introduction}
Collaboration is widely recognized as a powerful context for learning and problem solving \cite{nokes2019collaborative}. It has the potential to enhance not only group performance on the task at hand but also subsequent individual problem solving \cite{laughlin2008group, lippold2021gi}. However, simply working in groups is not always sufficient to produce meaningful gains \cite{do2025person}. Productive collaboration depends on the extent to which learners engage in high-quality dialogue that supports the generation, refinement, and integration of ideas \cite{barron2003smart, dillenbourg1999collaborative}.

In the present study, we measured cognitive engagement in collaborative discourse using complementary approaches based on trained human annotation and large language models (LLMs). Human annotators can identify theoretically meaningful distinctions in discourse through contextual interpretation and structured coding frameworks \cite{chi1997quantifying, krippendorff2018content}, whereas LLM-based methods offer strong potential for scalable analysis across large datasets \cite{gilardi2023chatgpt, tan2024large}. We evaluated the reliability of LLM-based approaches for characterizing cognitive engagement and supporting refinement of the engagement framework relative to trained human annotation.

\subsection{Cognitive Engagement in Collaborative Discourse}
A prominent theoretical framework for characterizing cognitive engagement is the ICAP model, which distinguishes Passive, Active, Constructive, and Interactive modes of engagement based on whether learners contribute information beyond what is already available and whether they coordinate their reasoning with others \cite{chi2009active, chi2014icap}. Passive engagement involves attending without overt contribution, whereas active engagement involves participation that supports the discussion without introducing new understanding such as expressing agreement or repeating what was already stated. Constructive engagement involves extending understanding by generating explanations, summaries, interpretations, or questions. Interactive engagement involves sustained exchanges in which learners build on one another’s reasoning. These modes form an ordered hierarchy of engagement, with interactive engagement associated with the most robust learning outcomes, followed by constructive, active, and passive engagement \cite{menekse2013differentiated}.

In collaborative problem-solving settings, important variation can arise within the same mode of participation, as learners’ contributions differ in their depth and quality of engagement. For example, within the passive category, remaining silent reflects a different level of involvement than providing brief acknowledgments (e.g., “yes” or “uh-huh”), which signal attention and help maintain coordination among group members \cite{schegloff1982discourse}. Questions also vary in how much they advance understanding: simple clarification or fact-checking questions reflect active engagement, whereas higher-level “how” or “why” questions extend the discussion and signal movement toward constructive engagement \cite{graesser1994question, king1992facilitating}. Similarly, contributions can range from partially developed ideas to well-supported explanations articulated with a clear rationale, with more elaborated explanations associated with greater learning \cite{lachner2025does}. Interactive engagement likewise varies in depth, depending on whether learners briefly respond to others’ ideas or engage in sustained exchanges that iteratively refine interpretations. Short sequences of constructive turn-taking reflect emerging coordinated reasoning, whereas extended back-and-forth exchanges reflect more sustained co-construction of shared understanding \cite{stahl2006computer}. 

To represent these gradations while preserving the theoretical ordering of engagement proposed in the ICAP framework, engagement in the present study was coded using a 7-point scale anchored to the four engagement modes. Introducing intermediate levels between adjacent categories allowed us to capture differences in participation when instances did not align cleanly with a single mode. This approach enables more fine-grained analyses of how engagement unfolds over time and provides a foundation for examining how subtle differences in participation may contribute to variation in learning outcomes during collaborative problem solving.

\subsection{Large Language Models for Text Analysis}

LLMs have increasingly been applied as scalable tools in psychological research, particularly for text-based labeling tasks in which models assign psychological categories to qualitative responses or discourse segments \cite{bermejo2025llms,lu2025systematicbiaslargelanguage}. For example, prior work has shown that a fine-tuned GPT-3 model can outperform traditional manual coding in capturing orchestration actions in collaborative learning scenarios, highlighting the potential of LLM-based coding in complex psychological and educational contexts \cite{amarasinghe2023gptcoding}. 



One capability of LLMs that has recently received particular attention is in-context learning (ICL) \cite{brown2020language}. This approach allows models to perform unseen tasks without retraining or updating weights by conditioning on a number of examples provided in the prompt. In some cases, ICL can achieve strong performance that approaches or even matches the performance of fine-tuning \cite{yin-etal-2024-deeper}. Despite its effectiveness, it also has important limitations. Previous research has shown that ICL is highly sensitive to prompt design, as the selection, ordering, and quality of examples may lead to various outcomes \cite{liu2021makesgoodincontextexamples}. Furthermore, standard ICL lacks explicit, systematic, and persistent mechanisms to support complex, multi-step annotation pipelines that require ongoing iteration, feedback, and reflection.

In contrast to the one-shot prompting paradigm of standard ICL, LLM-based agents support more interactive workflows that enable decision-making and multi-step reasoning \cite{xi2023risepotentiallargelanguage}. These agents can maintain state across turns, perform iterative reasoning, and take actions, which may make them better suited for complex labeling tasks and more complicated systems. Accordingly, recent work has explored the use of LLM agents to support annotation processes. For example, \textit{MEGAnno+} proposes a collaborative annotation framework in which humans and LLMs jointly generate reliable labels \cite{kim2024megannohumanllmcollaborativeannotation}. Similarly, recent work introduces a multi-agent system that coordinates task assignment, data labeling, and quality management within the annotation pipeline \cite{qin-etal-2025-crowdagent}. Motivated by these developments, the present study compares human annotation, ICL-based labeling, and LLM agent-based approaches for coding cognitive engagement in collaborative discourse.



\section{The Present Study}
The present study used an extended 7-point scale based on the ICAP framework to capture different levels of cognitive engagement during collaborative problem solving. Prior work has shown that extensions of the ICAP framework can help capture finer-grained variation in learners’ cognitive engagement in collaborative learning contexts \cite{vosniadou2023using}. Building on this line of work, trained coders in the present study annotated learners’ behaviors and iteratively refined category boundaries and coding criteria across two versions.

To compare human annotation with automated methods, we evaluated LLM-based labeling under both ICL and agent-based settings. LLM-based approaches can achieve relatively high accuracy using ICL and fine-tuning in cognitive research involving text analysis \cite{yin-etal-2024-deeper, amarasinghe2023gptcoding}. However, ICL relies on single-pass prompting, which may limit agreement with human annotation. Work on agentic LLM systems has explored interactive workflows to improve annotation efficiency, cost control, and coordination \cite{kim2024megannohumanllmcollaborativeannotation, qin-etal-2025-crowdagent}. Yet these approaches typically do not model iterative reflection and refinement of the coding framework in response to ambiguous cases, as human annotators do. To address this gap, we propose an agentic LLM system that assigns labels while identifying ambiguities and iteratively refining the coding framework. We conducted a comparative evaluation of three methods---human annotation, ICL-based labeling, and agent-based LLM annotation---to examine differences in annotation reliability and framework refinement. Our code is available at \url{https://github.com/HanlingJ/human-icl-llm-agent}.

\section{Method}

\subsection{Task and Dataset}
We measured cognitive engagement during collaborative problem solving by analyzing discourse from three-person groups solving a logical inference task in which ten letters (A–J) were randomly assigned numerical values (0–9) without replacement. Participants inferred these hidden assignments through iterative reasoning across multiple trials \cite{do2025person}. On each trial, groups proposed an arithmetic expression using letters and $+$ or $-$ signs (e.g., $A + B$), received the result in letter form (e.g., CE), and then guessed the value of one letter with immediate feedback. This structure enabled trial-level coding of individual cognitive engagement as groups progressed toward solving the full mapping.

Participants were undergraduate students at the authors’ institution. All participants provided informed consent, and their discussions were video-recorded and transcribed. The final dataset included 42 group conversations (averaging 11 minutes and six trials per group).

\subsection{Human Annotation of Cognitive Engagement}
Two trained annotators classified cognitive engagement in the collaborative discourse. We used an extended 7-point scale based on the ICAP framework to capture intermediate levels of engagement between the four original modes \cite{chi2014icap}. Annotators labeled participants’ cognitive engagement on each trial of the task, including both the expression and the subsequent guess.

The 7-point scale ranged from low to high engagement. Level 1 (Passive) reflected listening without overt contribution. Level 2 captured minimal responses while primarily listening (e.g., brief acknowledgments such as “uh-huh”). Level 3 (Active) included observable participation such as expressing agreement or repeating a peer’s statement. Level 4 (Initiating Constructive) represented emerging constructive engagement, such as asking higher-order “how” or “why” questions or introducing new input without elaborated reasoning. Level 5 (Constructive) involved proposing a new expression or guess with reasoning. The highest levels reflected sustained interactive engagement. Level 6 consisted of two-turn exchanges between constructive participants, and Level 7 consisted of three or more back-and-forth turns in which group members jointly developed or defended ideas. A turn was defined as all utterances produced by one speaker until another speaker began speaking.

Annotators independently coded group conversations and met regularly to resolve disagreements through discussion. This annotation process unfolded in three stages, reflecting the gradual refinement of the 7-point scale (Appendix~\ref{Criteria Versions}D). 
\begin{scales}
\item Stage 1 (Video data 1--10) included 306 labeling and reasoning samples. During this stage, the first author worked with the two annotators to establish and refine the rules and criteria for the 7-point Likert scale adapted from the ICAP framework. This version is referred to as \textbf{Criteria Version 1}.
The first author served as a mediator, helping the annotators resolve disagreements and reach consensus. 

\item Stage 2 (Video data 11--31) included 660 labeling and reasoning samples. The two annotators worked independently without mediator involvement, applying the criteria developed during Stage 1. They met after every two videos to discuss their annotations and resolve disagreements on their own, without additional guidance. 

\item Stage 3 (Video data 32--42) included 315 labeling and reasoning samples. Following Stage 2, the first author and two coders collaboratively developed \textbf{Criteria Version 2}, with more detailed coding criteria. All videos in Stage 3 were coded using this revised framework. Annotators were asked to reference specific codebook criteria to justify their classifications of each behavior.
\end{scales}

\subsection{LLM Model Selection} Two models were used in this study: GPT-4o \cite{openai2024gpt4ocard} and GPT-5.2 \cite{singh2025openaigpt5card}. Given the complexity of the task, we selected relatively advanced models, as prior work suggests they demonstrate stronger reasoning and produce more structured outputs.

\subsection{In-context Learning}

ICL was used to adapt the LLM to the cognitive engagement annotation task using a few examples provided in the prompt, without retraining or parameter updates. This approach allowed us to compare machine-based annotation with human annotators and examine whether different versions of the engagement criteria produced similar improvements in agreement across methods.


\textbf{Data Splitting.}
For ICL, our trained human annotators segmented the video transcripts into trials by identifying the start and end of each trial. These trial-level segments were then used as inputs to the LLMs.

\begin{algorithm}[!htbp]
\caption{ICL for Annotation}
\label{alg:icl}
\DontPrintSemicolon

\KwIn{
Criteria $C$; 
 labeled examples $\{(x_i, y_i,r_i)\}$; 
 data $x$.\\
}
\KwOut{Predicted label $\hat{y}$, reasoning $\hat{r}$ .}

\BlankLine
\textbf{1. Build prompt.}\;
$P \leftarrow I \,\|\, C$ \tcp*{instructions and criteria}
\ForEach{example $(x_i, y_i,r_i)$}{
  $P \leftarrow P \,\|\, (x_i, y_i,r_i)$ \tcp*{add ICL examples}
}
$P \leftarrow P \,\|\, x$ \tcp*{add the data sample}

\BlankLine
\textbf{2. Ask the LLM.}\;
$(\hat{y}, \hat{r}) \leftarrow \mathrm{LLM}\, f_{\theta}(P)$ \tcp*{LLM generates the answer}

\end{algorithm}







\textbf{Prompt and Examples.} 
Our prompt followed this format: task description, output rules, the 7-point scale, labeled examples and test samples, and a final emphasis on the output rules (Appendix~\ref{ICL_prompt}A). The ICL examples were drawn from the human coders’ consensus codebook, in which all disagreements were resolved and final labels were jointly agreed upon by both annotators. These annotations were used as the ground truth for constructing ICL examples and evaluating model performance. Examples were randomly sampled from a separate transcript that was strictly disjoint from the test transcript, ensuring no overlap between the example and test data.





\subsection{Self-Reflective Agent}
Inspired by the iterative refinement process of the two human coders, we designed an LLM-based agent that performed both annotation and framework refinement. After labeling each data sample, the agent performed reflection to identify ambiguities in the current framework and made decisions about whether to modify, add, delete, hold, or merge classes before proceeding to subsequent samples (Appendix~\ref{agentactions}C).

The agent first labeled each sample, then identified weaknesses in the current framework and generated reflections and revision suggestions (Appendix~\ref{WandAAnalysis}B). At each iteration, the agent retained short-term memory of the three most recently labeled samples and the three most recent versions of the criteria, which served as context for reflection and labeling the next sample. This iterative memory enabled progressive self-refinement beyond a single-prompt labeling setup.


\begin{figure}[h]
    \centering
    \includegraphics[width=1.0\linewidth]{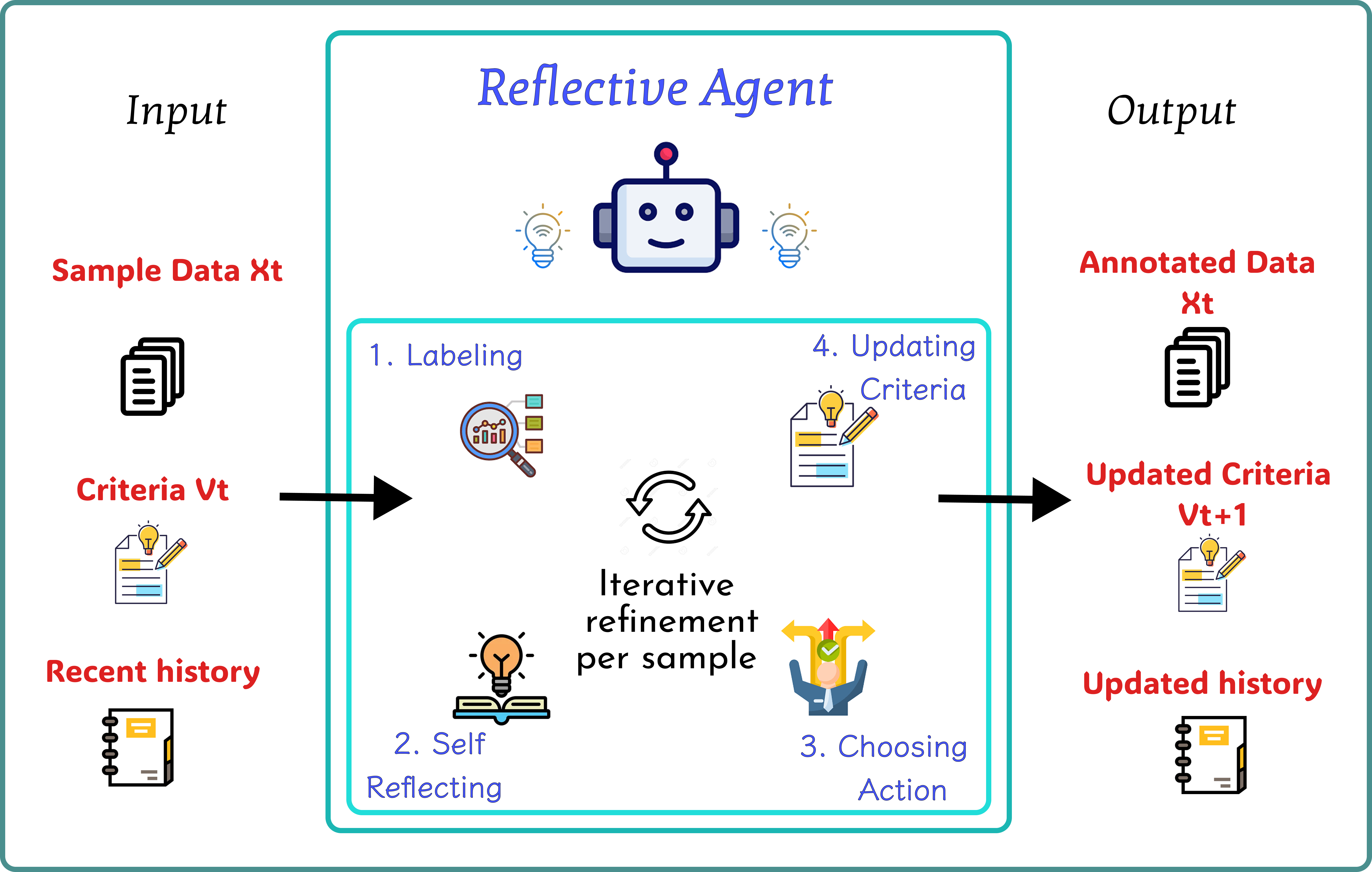}
    \caption{Reflective LLM Agent System With Memory Workflow.}
    \label{fig:placeholder}
\end{figure}




  
  

  
  

  

  

\begin{algorithm}[h]
\caption{LLM-based Agent with Reflection and Criteria Update}
\label{alg:agentic}
\DontPrintSemicolon

\KwIn{
Initial criteria $C^{(0)}$; 
data $\{x_t\}_{t=1}^{T}$.
}

\KwOut{
Annotations $\{(x_t, \hat{y}_t, \hat{r}_t)\}$;
Updated criteria $C^{(\text{final})}$.
}

\BlankLine
\For{$t \leftarrow 1$ \KwTo $T$}{
  \BlankLine
  \textbf{1. Label current sample}\;
  
  $P_{\text{label}} \leftarrow$ instructions, criteria $C$, history $H_{\text{buf}}$ with most recent $K$ samples, and sample $x_t$\;
  
  $(\hat{y}_t, \hat{r}_t) \leftarrow \mathrm{LLM}\, f_{\theta}(P_{\text{label}})$

  \BlankLine
  \textbf{2. Reflect on criteria}\;
  
  Build Prompt $P_{\text{ref}}$ from $C$, $H_{\text{buf}}$, and $(x_t, \hat{y}_t, \hat{r}_t)$\;
  
  $\Delta \leftarrow \mathrm{LLM}\, f_{\theta}(P_{\text{ref}})$

  \BlankLine
  \textbf{3. Update criteria}\;
  
  $C \leftarrow \textsc{ApplyChanges}(C, \Delta)$\;

  \BlankLine
  \textbf{4. Update history}\;
  
  Append $(x_t, \hat{y}_t, \hat{r}_t)$ to $H_{\text{buf}}$\;
}

\KwRet{
All annotations, $H$ history with all versions of the criteria and reflection, the final criteria $C^{(\text{final})} \leftarrow C$.
}
\end{algorithm}

 \section{Results}
\subsection{Interrater Reliability}
We used three agreement coefficients: quadratic weighted kappa (QWK) \cite{fleiss1973equivalence}, Krippendorff’s $\alpha$ \cite{hayes2007answering}, and Randolph’s $\kappa$ \cite{randolph2005free}. QWK captures chance-corrected agreement on the extended 7-point scale. Krippendorff’s $\alpha$ is a standard reliability index in content analysis. Randolph’s $\kappa$ is a free-marginal reliability coefficient that is less sensitive to class imbalance than fixed-marginal kappa measures. Together, these metrics provide converging evidence for the reliability analyses.

We compared the original ICAP framework with the extended 7-point scale using zero-shot learning (Table~\ref{tab:baseline}). Interrater reliability between the two models increased substantially when using the 7-point scale ($\Delta$QWK = 0.261, $\Delta$Randolph’s $\kappa$ = 0.427, $\Delta$Krippendorff’s $\alpha$ = 0.352), suggesting that the extended framework supports more consistent annotation. Additional results with the 7-point scale across human annotators, ICL, and LLM agents are reported in Table~\ref{tab:reliability}.

\begin{table}[H]
\centering
\caption{Interrater Reliability Between GPT-5.2 and GPT-4o With Zero-Shot Labeling Across Two ICAP Versions.}

\begin{tabular}{lcc}
\hline
\textbf{Metric} & \textbf{7-point Scale} & \textbf{Original ICAP} \\
\hline
QWK & 0.874 & 0.613 \\
Randolph's $\kappa$ & 0.609 & 0.182 \\
Krippendorff's $\alpha$ & 0.839 & 0.487 \\
\hline
\end{tabular}
\label{tab:baseline}
\end{table}


\textbf{Human–Human and Agent–Agent Reliability.}
Interrater agreement between the two human annotators ($H^1H^2$) was high in Stage~1 (QWK = 0.906; Krippendorff's $\alpha$ = 0.873) and increased further as the framework was refined, reaching near-perfect agreement by Stage~3 (QWK = 0.998; $\alpha$ = 0.999). Overall, reliability between the two human coders was very high (QWK = 0.974; Randolph's $\kappa$ = 0.894). 

Agreement between the two LLM agents ($A^1A^2$) was substantial but lower than human–human agreement. During the refinement process, the agents reached QWK = 0.841. After applying their separately refined frameworks under zero-shot prompting, agreement ranged from QWK = 0.655 to 0.790, suggesting variability in annotation consistency across foundation models. More advanced models may support more stable and reliable framework refinement.

\newcommand{\best}[1]{\cellcolor{blue!15}\textbf{#1}}
\newcommand{\worst}[1]{\cellcolor{red!15}\textbf{#1}}

\begin{table*}[h]
\caption{Interrater Reliability Across Human Coders, ICL and LLM Agents throughout Criteria Refinement Process}
\label{tab:reliability}
\vskip 0.10in

\resizebox{\textwidth}{!}{%
\begin{tabular}{p{0.12\textwidth}p{0.11\textwidth}p{0.10\textwidth}p{0.25\textwidth}ccc}
\toprule
Agreement between & ICL setting & Model & Condition &
QWK & Randolph's-$\kappa$ & Krippendorff's-$\alpha$ \\
\midrule

\multirow{4}{*}{$H^{1}H^{2}$}

  & N/A           &           & Stage 1  & \worst{0.906} & \worst{0.705} & \worst{0.873} \\
  &            &            & Stage 2  & 0.986         & 0.937         & 0.981 \\
  &            &            & Stage 3  & \best{0.998}  & \best{0.986}  & \best{0.999} \\
  &            &            & Overall  & 0.974         & 0.894         & 0.966 \\
\midrule

\multirow{1}{*}{$A^{1}A^{2}$}
  & N/A           &           & Refinement Stage 1           & 0.841  & 0.600  & 0.823 \\

\midrule

\multirow{16}{*}{$H^{\star}M$}
& ICL       & GPT-5.2 & Cri-V1 Stage 1,2   & \best{0.655} & 0.354 & \best{0.641} \\
&           &         & Cri-V2 Stage 3   & \worst{0.593}        & 0.351         & \worst{0.575} \\
&           &         & Cri-V1 Overall   & 0.631        & \worst{0.327} & 0.622 \\
&           &         & Cri-V2 Overall   & 0.635        & \best{0.369}  & 0.622 \\
\cmidrule(lr){2-7}
& Zero-shot & GPT-5.2 & Cri-V1 Stage 1,2   & \best{0.649} & 0.315         & \best{0.624} \\
&           &         & Cri-V2 Stage 3   & \worst{0.518}& \worst{0.255} & \worst{0.580 }\\
&           &         & Cri-V1 Overall   & 0.629        & 0.300         & 0.613 \\
&           &         & Cri-V2 Overall   & 0.608        & \best{0.353}  & 0.594 \\
\cmidrule(lr){2-7}
& ICL       & GPT-4o  & Cri-V1 Stage 1,2   & \best{0.631} & 0.340 & \best{0.616} \\
&           &         & Cri-V2 Stage 3   & \worst{0.585}        & \best{0.417}  & 0.589 \\
&           &         & Cri-V1 Overall   & 0.607        & \worst{0.323} & 0.594 \\
&           &         & Cri-V2 Overall   & 0.589& 0.367         & \worst{0.569} \\
\cmidrule(lr){2-7}
& Zero-shot & GPT-4o  & Cri-V1 Stage 1,2   & \best{0.640} & 0.323         & \best{0.626} \\
&           &         & Cri-V2 Stage 3   & \worst{0.523}& \best{0.332}  & \worst{0.509} \\
&           &         & Cri-V1 Overall   & 0.624        & \worst{0.315} & 0.617 \\
&           &         & Cri-V2 Overall   & 0.576        & 0.341         & 0.559 \\
\midrule

\multirow{6}{*}{$M^1M^2$}
  & \multirow{2}{*}{ICL}
  &            & Cri-V1 Overall & 0.882         & \worst{0.541} & 0.847 \\
  &            &            & Cri-V2 Overall & \worst{0.851} & 0.594         & \worst{0.810} \\
  & \multirow{2}{*}{Zero-shot}
  &            & Cri-V1 Overall & \best{0.894}  & 0.547         & \best{0.853} \\
  &            &            & Cri-V2 Overall & 0.874         & \best{0.609}  & 0.839 \\
\cmidrule(lr){2-7}
  & Zero-shot            &            &  Cri-$A^{1}$ overall & \worst{0.655} & \worst{0.406} & \worst{0.636} \\
  &            &            & Cri-$A^{2}$ overall  & 0.790         & 0.588         & 0.762 \\
\bottomrule
\end{tabular}%
}
\vspace{0.5em}
\begin{tablenotes}
\footnotesize
\item \textit{Note.} $H^1$ and $H^2$ denote the human coders. $A^1$ and $A^2$ denote the GPT-4o and GPT-5.2–based agents. “Cri-V1” and “Cri-V2” refer to the human-refined criteria versions, whereas “Cri-$A^{1}$” and “Cri-$A^{2}$” refer to the criteria refined by the GPT-4o and GPT-5.2 agents. $H^{\star}M$ indicates agreement between the human consensus annotations and an LLM, whereas $M^1M^2$ indicates GPT-4o vs. GPT-5.2 agreement. “Stage” refers to phases of human annotation, and “Overall” aggregates results across all videos and stages. Purple represents the highest agreement values under each condition, and red represents the lowest.
\end{tablenotes}
\end{table*}





\textbf{Human--Machine Reliability.}
For machine-based annotation using GPT-5.2 with ICL prompting, human–machine QWK values ranged from 0.593 to 0.655 across stages and criteria versions, with Krippendorff’s $\alpha$ around 0.575–0.641. The human-refined framework produced only small changes in reliability from Criteria Version~1 to Version~2 (overall $\Delta$QWK $\approx 0.01$–0.04), in contrast to the larger gains observed for human–human agreement. A similar pattern held for GPT-4o, with overall QWK remaining around 0.585–0.631. Differences between ICL and zero-shot prompting were also modest and sometimes favored zero-shot, indicating that few-shot prompting with human examples did not consistently improve human–machine reliability on this task.

\textbf{Machine--Machine Reliability.}
Machine–machine reliability was higher than human–machine reliability but remained lower than human–human agreement. Under the ICL setting, QWK was 0.882 for Criteria Version~1 and 0.851 for Version~2, with corresponding Krippendorff’s $\alpha$ values of 0.847 and 0.810, respectively. Zero-shot prompting using human-refined framework produced slightly higher agreement (QWK = 0.894 and 0.874), suggesting that in-context examples did not improve annotation consistency once a clear framework was provided. Overall, agreement remained stable across prompting conditions, with only modest reliability gains from the human-refined framework.

\subsection{Human Annotation Disagreements and Framework Refinement}

\begin{figure}[h]
  \centering
  \includegraphics[width=0.9\linewidth]{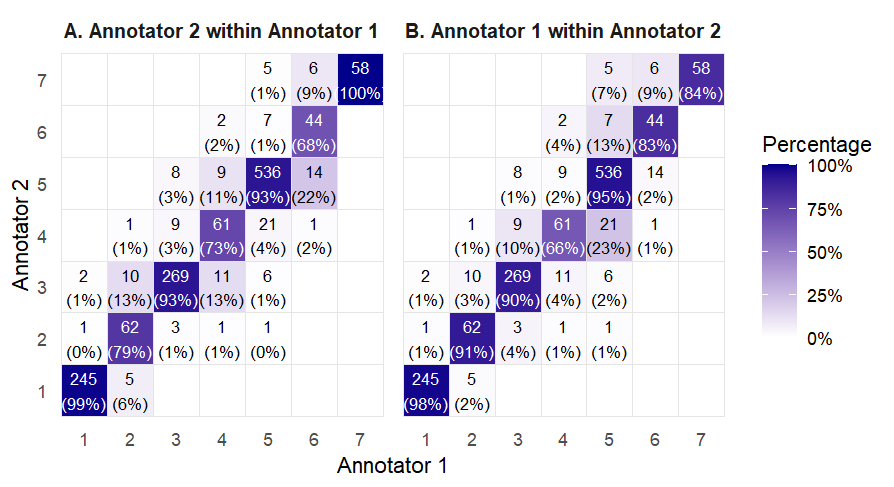}
  \caption{Confusion Matrix Between Human Annotators. Note: Panel A shows how Annotator 2 classified items within each Annotator 1 level, and Panel B shows how Annotator 1 classified items within each Annotator 2 level.}
  \label{fig:confusionmatrix}
\end{figure}

Human disagreements occurred most frequently between Levels 4 and 5 (24\% of all disagreement pairs), followed by Levels 5 and 6 (17\%) and Levels 3 and 4 (16\%), although this pattern may partly reflect the higher frequency of Level 5 behaviors. The confusion matrices further showed bidirectional confusion across levels (Figure~\ref{fig:confusionmatrix}). When one annotator coded Level 4, the other coded Level 5 in 11--23\% of cases. A similar pattern was observed between Levels 5 and 6, with 13--22\% of Level 6 ratings coded as Level 5 by the other annotator. Together, these findings indicate greater ambiguity in the mid-level categories than at the endpoints of the scale.

Consistent with this pattern of disagreement, human annotators made the most revisions to the mid-level categories, adding more detailed criteria to distinguish among Levels 3--5 (Appendix~\ref{Criteria Versions}D). Across the agent refinement process, Levels 3--5 likewise underwent more modifications than the other levels (Figure~\ref{fig_action}), mirroring the refinement trajectory observed for human coders. Although agents could take multiple types of refinement actions, they only modified existing levels and did not add, merge, or remove any categories, preserving the seven-level structure throughout refinement.

\begin{figure}[h]
    \centering
    \includegraphics[width=0.45\linewidth]{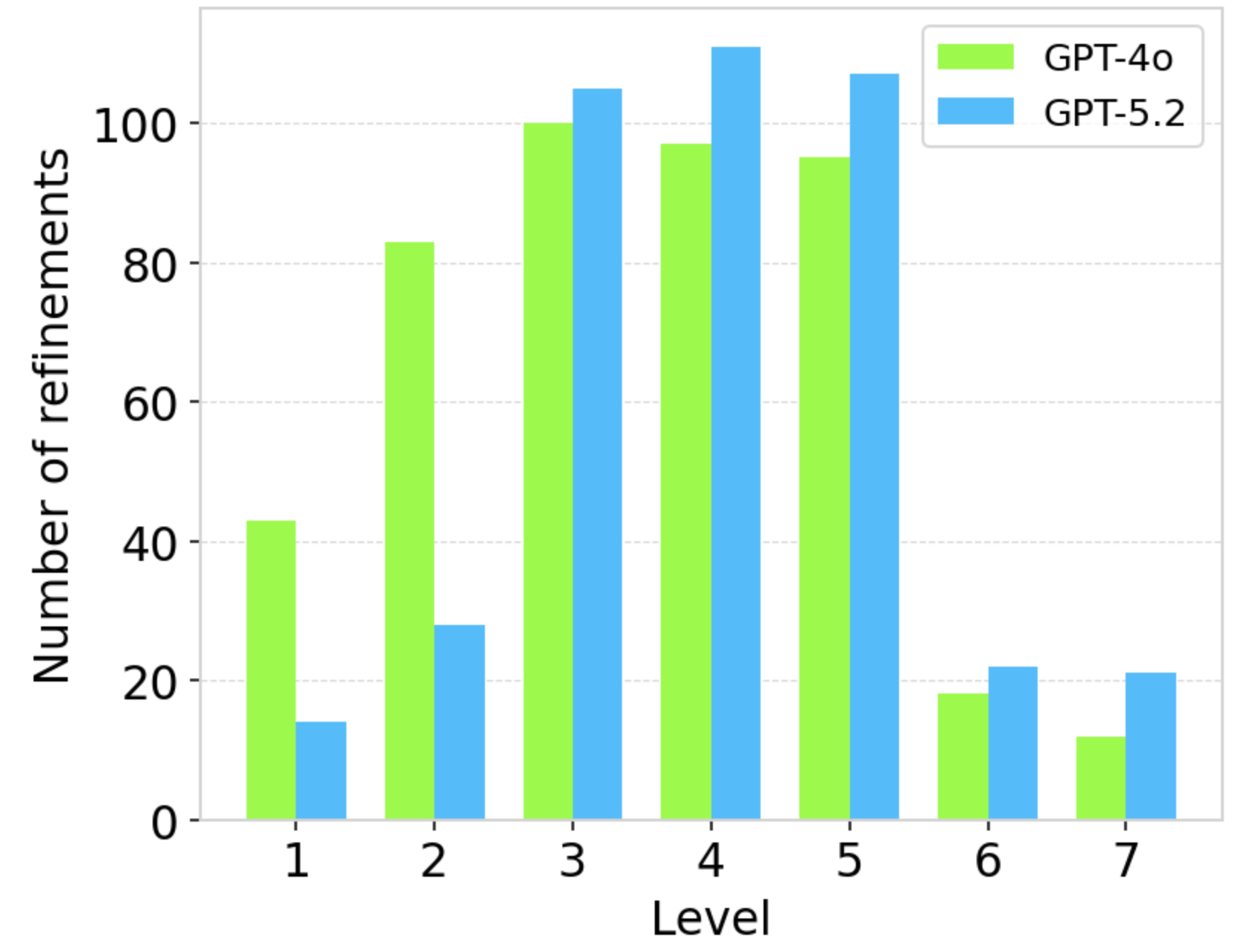}
    \label{fig_action}
    \centering
    \includegraphics[width=0.45\linewidth]{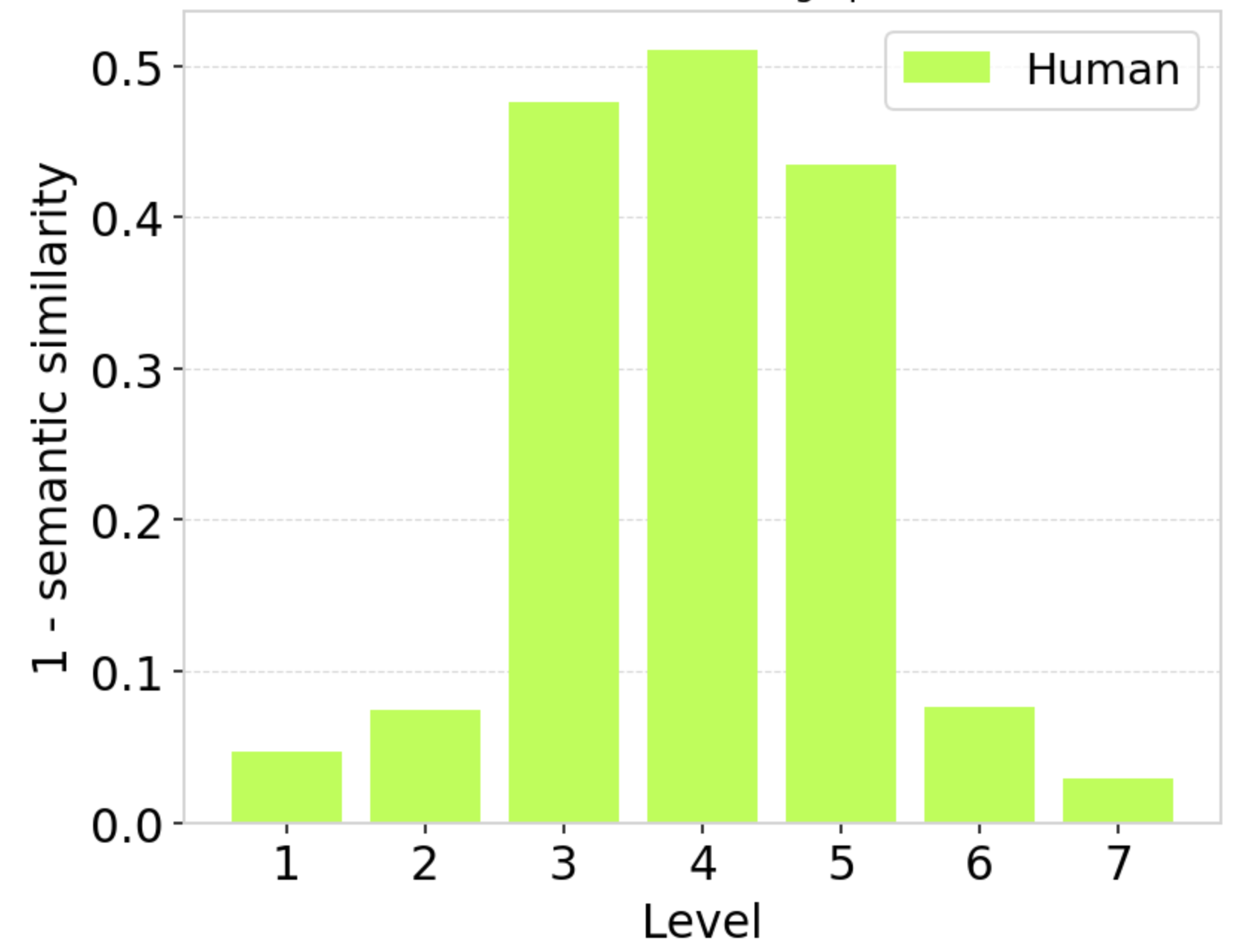}
    \caption{Agent Automated Refinement and Human Criteria Updates. Note: The left plot shows refinement counts per engagement level for the GPT-5.2 and GPT-4o agents, while the right plot shows $(1-\mathrm{CosSim})$ between the Criteria Version 1 and the Criteria Version 2 for each engagement level.}
    \label{fig_action}
\end{figure}

\section{Discussion}
The present study highlights both the promise of LLM-based agent annotation and important differences in how engagement coding frameworks operate for human versus LLM-based annotators. Human coders showed consistently high interrater reliability across the three stages of annotation, indicating that continued training and clarification of category definitions improved agreement. LLM-based annotation also showed moderate to relatively high interrater reliability in machine–machine and agent–agent agreement. However, agreement between human and LLM-based annotations remained more moderate, indicating that consistency within LLM-based annotation does not necessarily translate into closer alignment with human coding. Furthermore, applying the same human-refined criteria produced only small and sometimes inconsistent changes in reliability for LLM annotators across annotation stages, suggesting that refinements optimized for human interpretation do not directly translate into comparable improvements for models through modification of the task instruction in the prompt alone.

Our results also indicate that ICL does not necessarily enhance annotation consistency for LLMs in this setting. Across criteria versions and model configurations, ICL and zero-shot prompting produced similar levels of agreement, and in several cases zero-shot prompting slightly outperformed ICL. This pattern suggests that few-shot examples derived from human annotations may not provide a stronger signal than clear task instructions alone for complex, theory-based engagement coding tasks.

When comparing human-developed criteria with agent-refined frameworks, we found that both supported substantial agreement between LLMs, though with greater variability under agent refinement. Cross-model agreement remained relatively high under human-developed criteria (QWK $\approx 0.85$--$0.89$), but ranged more widely under agent-refined criteria (QWK $\approx 0.66$--$0.79$). Notably, both human annotators and agents concentrated revisions on the mid-range engagement levels, particularly Levels 3--5 (Figure~\ref{fig_action}). This pattern suggests greater ambiguity in the mid-level categories, whereas the lower and upper levels remained relatively stable. These findings suggest that refining mid-level category boundaries may improve extended ICAP frameworks in future work.

\section{Limitations and Future Directions}
The first limitation concerns differences in the information available to human versus LLM-based annotators. Human coders had access to both video recordings and transcripts, whereas LLMs were restricted to text-only inputs. Although human annotators were instructed to focus on explicit verbal behaviors, they were naturally exposed to cues such as tone, timing, and nonverbal behavior while watching the videos. The restriction of LLMs to text-only inputs was due to the high computational cost of processing long video data with large vision–language models and may have reduced the sensitivity of LLM-based annotation to multimodal aspects of engagement. A second limitation concerns differences in how framework refinement unfolded for human annotators and LLM agents. Human refinement occurred across three stages, whereas LLM agents refined the framework at each iteration. Additionally, human annotators frequently discussed disagreements during coding, whereas agent-based refinement relied solely on each agent’s own analysis. The high interrater reliability observed for human coders may partly reflect these opportunities for discussion. Future work should adopt more comparable refinement procedures and include independent annotators outside the consensus process to better evaluate their relative contributions.

Future work could also extend this research in two directions to further enhance scalability and annotation support. First, it is important to examine whether LLMs can be fine-tuned using the human-refined framework and annotated data, to assess whether models can acquire implicit knowledge not fully captured by the explicit criteria. Second, future research could explore LLM agent systems from two perspectives: hybrid systems in which human coders and LLM agents collaborate, and multi-agent systems in which multiple agents coordinate without human involvement. Both approaches may help refine the coding framework, identify challenging cases, and support annotation and quality control. A key question is whether LLMs can make reliable autonomous decisions and align with human judgments, which has important implications for redefining the roles of LLMs and human annotators in text analysis. Ultimately, these approaches may improve reliability while reducing the workload of human annotators.

\section{Acknowledgments}
This work was supported by the National Science Foundation (NSF) under Grant No.\ 2428640, \textit{GCR: Towards a Convergent Understanding of the Dynamics of Uncertainty in Individuals and Groups with a Focus on STEM Education}. In this work, Lan Anh Do contributed to conceptualization, development of the human-refined framework, human-annotation training and supervision, data collection, data curation, formal analysis, and writing (original draft, review, and editing). Hanling Jiang contributed to the LLM-based methodology, software development and validation, experimental implementation and evaluation, data curation, formal analysis, and writing (original draft, review, and editing). Ayanna Thomas and Shuchin Aeron contributed to conceptualization, supervision, and writing (review and editing). AI tools were used for code development and to assist with proofreading.





\bibliographystyle{apacite}

\setlength{\bibleftmargin}{.125in}
\setlength{\bibindent}{-\bibleftmargin}

\bibliography{CogSci_Template}

\newpage
\newpage
\appendix


\section{Appendix A: In-Context Learning}
\label{ICL_prompt}

\subsection{Prompt}

The in-context learning prompt consists of the task description, output rules, the 7-point framework, labeled examples, test samples, and a final emphasis on the output rules.

\noindent\textbf{Example:}

\begin{quote}
\small
You will be given a classification task involving student engagement in a group discussion while playing the letters-to-number game.
Your goal is to analyze each conversation and, for each subject, assign a 7-point Likert scale score and provide a brief reasoning based on their statements.
You will first see a few examples. Then, you will be given a new task to label.

\medskip
\noindent Please follow these rules:
\begin{itemize}[leftmargin=*, itemsep=0pt, topsep=2pt]
    \item For each \textbf{Subject}, assign a Likert scale score from 1 to 7 and provide a concise reasoning for your choice.
    \item Follow the structure in the example.
    \item Only provide labels and reasoning for trial(s) shown in TASK. \textbf{Do not} make up or add any trial that is not included.
\end{itemize}

\medskip
\noindent\textbf{[Insert examples and test samples here.]}

\medskip
\noindent Please follow these rules:
\begin{itemize}[leftmargin=*, itemsep=0pt, topsep=2pt]
    \item For each \textbf{Subject}, assign a Likert scale score from 1 to 7 and provide a concise reasoning for your choice.
    \item Follow the structure in the example.
    \item Only provide labels and reasoning for trial(s) shown in TASK. \textbf{Do not} make up or add any trial that is not included.
\end{itemize}

\noindent For each subject:
\begin{enumerate}[leftmargin=*, itemsep=0pt, topsep=2pt]
    \item Assign a Likert score according to the criteria.
    \item Provide short reasoning.
    \item Output your annotation following the format below.
\end{enumerate}

\medskip
\noindent\texttt{Trial n expression}\\
\texttt{subject\_1 (Subject ID): label = x, reasoning = one line}\\
\texttt{subject\_2 (Subject ID): label = x, reasoning = one line}\\
\texttt{subject\_3 (Subject ID): label = x, reasoning = one line}

\medskip
\noindent\texttt{Trial n guess}\\
\texttt{subject\_1 (Subject ID): label = x, reasoning = one line}\\
\texttt{subject\_2 (Subject ID): label = x, reasoning = one line}\\
\texttt{subject\_3 (Subject ID): label = x, reasoning = one line}

\medskip
\noindent Again, only provide labels and reasoning for trial(s) shown in TASK. \textbf{Do not} make up or add any trial that is not included.
\end{quote}

\section{Appendix B: Example of Agent Reflection}
\label{WandAAnalysis}
\textbf{Reflection}
\newline
The current codebook leaves gaps that caused interpretive decisions: \newline
(a) proposals without justification are not explicitly placed (we treated them as 4)\newline
(b) weak heuristics blur 4 vs 5\newline
(c) 6–7 rely on vague 'turn-taking' language that can mislabel parallel contributions as interactive\newline
(d) the 1–3 boundary between silence, backchannel, and simple task talk is underspecified. The proposed updates supply clear decision rules and examples aligned with the observed utterances (e.g., 'random guess' as 4; 'mid-range' without why as 4; constraint-based reasoning as 5) and make interaction levels reliably detectable by requiring explicit builds across turns.

\section{Appendix C: Actions Assigned to the Agent for Framework Refinement}
\label{agentactions}

\begin{itemize}
    \item \textbf{Modify}: refine an existing scale definition.
    \item \textbf{Add}: introduce a new scale level.
    \item \textbf{Merge}: merge multiple scales into a target scale.
    \item \textbf{Delete}: remove a scale entirely.
    \item \textbf{Rename}: reindex a scale, such as shifting numbers, or rename keys.
    \item \textbf{Hold}: no changes.
\end{itemize}

Note: If no action was detected, \textit{hold} was automatically selected. Although the agent was allowed to take different actions, in our experiments it consistently chose only to modify one or more of the seven levels and never proposed adding, merging, or deleting any level.

\section{Appendix D: Criteria Versions}
\label{Criteria Versions}
\subsection{Criteria Version 1}
\begin{quote}
\small\ttfamily
{\footnotesize
\begin{scales}
  \item[Level 1 (Passive):]
        The student does not say anything.
  \item[Level 2 (Between Passive and Active):]
        Minimal responses like 'uh-huh'.
  \item[Level 3 (Active):]
        Agrees, repeats others, asks simple clarifications.
  \item[Level 4 (Between Active and Constructive):]
        Adds reasoning or asks 'why'/'how'.
  \item[Level 5 (Constructive):]
        Suggests expressions or guesses with reasoning, explains situation.
  \item[Level 6 (Between Constructive and Interactive):]
        Two turn-taking between two constructive participants.
  \item[Level 7 (Interactive):]
        Three or more turn-taking between two constructive participants.
\end{scales}
}
\end{quote}

\subsection{Criteria Version 2}
\begin{quote}
\small\ttfamily
{\footnotesize
\begin{scales}
  \item[Level 1 (Passive):]
        Not saying anything.
  \item[Level 2 (Between Passive and Active):]
        Has a minimal response, such as "uh-huh."
  \item[Level 3 (Active):]
        Agreed and/or repeated what others said.\\
        Asked simple questions for basic clarification (e.g., "What did you just say?" or whether they could divide the letters).\\
        Simple correction (e.g., when all members agreed that C was 6, but the host mistakenly entered 7, and someone corrected her).\\
        Expressions and guesses based on others’ suggestions and kept track of the group’s progress by taking notes, without contributing new information beyond what others already knew.
  \item[Level 4 (Between Active and Constructive):]
        Agreed and added their own reasoning.\\
        Asked advanced questions that led to constructive information, such as "how" and "why" types.\\
        Provided some new input (e.g., an alternative expression; or says the expression should go up to J) but did not elaborate the reasoning.\\
        Attempted to provide new input, but failed to complete their reasoning.\\
        Disagreed without elaborating the reasoning.
  \item[Level 5 (Constructive):]
        Suggested a new expression or guess with reasoning (or implicitly conveys reasoning that is understood by members). This includes suggestions such as making a random guess.\\
        Answered questions.\\
        Explained or summarized the situation to a member.\\
        Identified possible values or ruled out impossible values for a specific letter or number.\\
        Disagreed with reasoning.
  \item[Level 6 (Between Constructive and Interactive):]
        Two turn-taking between two constructive participants.
  \item[Level 7 (Interactive):]
        Three or more turn-taking between two constructive participants.
\end{scales}}
\end{quote}

\end{document}